\documentclass[aps,pre,twocolumn,superscriptaddress]{revtex4}  
\usepackage{graphicx,epsfig}  
\usepackage{dcolumn}   
\usepackage{bm}        
\usepackage{amssymb}   
\usepackage{color}
\usepackage{mathtools}
\usepackage{esvect}
\usepackage{float}
\begin{document}
\title{ Self-excited converging shock structure in a complex plasma medium }%
\author{Garima Arora}%
\email{garimagarora@gmail.com}
\affiliation{Institute of Plasma Physics of the Czech Academy of Sciences, Prague, Czech Republic}%
 \author{Srimanta Maity}
 \email{srimantamaity96@gmail.com}
 \affiliation{ELI Beamlines Facility, The Extreme Light Infrastructure ERIC, Za Radnicí 835, 25241 Dolní Břežany, Czech Republic}%

\begin{abstract} 
 We report the study of a self-excited converging shock structure observed in a complex plasma medium for the first time. A high-density dust cloud of melamine formaldehyde particles is created and horizontally confined by a circular ring in a DC glow discharge plasma at a particular discharge voltage and pressure. Later on, as the discharge voltage is increased, a circular density crest is spontaneously generated around the outer boundary of the dust cloud. This nonlinear density structure is seen to propagate inward towards the center of the dust cloud. The properties of the excited structure are analyzed and found to follow the characteristics of a converging shock structure. A three dimension molecular dynamics (MD) simulation has also been performed in which a stable dust cloud is formed and levitated by the balance of forces due to gravity and an external electric field mimicking the cathode sheath electric field in the experiment. Particles are also horizontally confined by an external electric field, representing the sheath electric field of the circular ring present in the experiment. A circular shock structure has been excited by applying an external perturbation in the horizontal electric field around the outer boundary of the dust cloud. The characteristic properties of the shock are analyzed in the simulation and qualitatively compared with the experimental findings. This study is not only of fundamental interest but has many implications concerning the study of converging shock waves excited in other media for various potential applications.
\end{abstract}
\maketitle
\section{Introduction}\label{sec:intro}
Self-excited structures are the sources of fingerprints in determining the underlying physical phenomena in a medium. For instance, the localization or self-organization is studied as auto-waves of plastic strain \cite{vasiliev2012autowave}. Observing hydro-thermal waves or thermal patterns at the liquid-vapor interface reveals the mechanism of heat transfer and energy transport \cite{sefiane2013thermal}. The self-excitation of upstream magneto-hydrodynamics waves in the region of Earth's bow shock is one of the outstanding physical phenomena in solar-terrestrial physics concerning nonlinear wave-particle interactions and particle accelerations \cite{lee1982coupled}. Even in a laboratory plasma, the self-excited ionization waves (striations) are the sources of nonlinear and turbulent effects \cite{krasa1974evolution}. Hence, a detailed study of self-excited dynamics is necessary to understand the hidden phenomena revealing its fundamental origin. Our present study is devoted to the self-generated structure in a complex plasma medium. In particular, we have investigated the characteristic properties of a self-excited converging shock structure observed in a complex plasma experiment. 
 
Complex plasmas \cite{morfill2009complex,fortov2005complex} are low-temperature discharge plasmas with embedded submicron/micron-sized particles. Once introduced in the plasma, particles become negatively charged as a larger number of electrons, being lighter than ions, stick to their surfaces. Because of the acquired high charge,  their mutual interactions become more substantial concerning their average thermal energy and demonstrate the formation of the crystalline state in various laboratory experiments \cite{thomas1994plasma,hariprasad2018experimental}. Moreover, it exhibits phase transitions from crystalline to liquid form due to various instabilities \cite{maity2022parametric,joyce2002instability,hariprasad2020experimental} and complex processes in the medium \cite{vasilieva2021laser}. The study on the structural phase transition of a complex plasma crystal has been reported in a previous study \cite{maity2019molecular}. It has also been proven a versatile medium to study collective excitations such as linear \cite{barkan1995laboratory} and nonlinear waves \cite{bandyopadhyay2008experimental}, modes \cite{yaroshenko2005coupled}, structures \cite{samsonov2004shock,heinrich2009laboratory,maity2018interplay}, voids \cite{dahiya2002evolution}, and vortexes \cite{choudhary2017experimental} in various equilibrium state regimes. Additionally, the investigation of spectacular phenomena such as cluster formation \cite{maity2020dynamical, deshwal2022chaotic}, generation of voids \cite{dahiya2002evolution}, and vortexes \cite{choudhary2017experimental} in a dusty plasma medium has also been reported in the literature. Although the medium has various complexities, it has been studied well, as the particles can easily be tracked and observed in the laboratory. However, it still poses challenges due to the dynamic behavior of plasma electrons and ions in the background \cite{arora2019micro,arora2018dust} and, thus, has remained an important topic of fundamental research. 
 
Various studies have been reported on the collective excitation of driven structures in a dusty plasma medium, such as longitudinal dust acoustic waves \cite{barkan1995laboratory}, transverse shear waves \cite{pramanik2002experimental}, solitons \cite{bandyopadhyay2008experimental, kumar2017observation}, precursor solitons \cite{arora2019effect}, pinned solitons \cite{arora2021experimental}, and shocks \cite{samsonov2004shock,heinrich2009laboratory,arora2020excitation}. Recently, in a simulation study, the generation of a new wave called transverse surface wave in a monolayer complex plasma medium has also been reported \cite{maity2023amplitude}. Besides the controlled excitations of collective waves, there are reports on self-excited structures in the laboratory as well as experiments performed in microgravity. In the parabolic flight experiments, Schwabe \textit{et. al. } \cite{schwabe2007highly} observed the self-excited dust density waves propagating in the direction of ion drift. The self-excited dust acoustic waves and their breaking have been studied by Teng\textit{et. al.} \cite{teng2009wave}. Heinrich \textit{et al.} \cite{heinrich2009laboratory} reported the observation of self-excited dust acoustics shocks and two colliding shock structures in a DC glow discharge dusty plasma. The spontaneous rotation of dust torus due to gradient in ion drag force has also been studied by Manjit \textit{et. al.} \cite{kaur2015observation}. Recently, Choudhary \textit{et. al.} \cite{choudhary2017experimental} reported the observation of self-excited vortices in the absence of a magnetic field. Dahiya \textit{et. al.} \cite{dahiya2002evolution} studied the transition of a circular symmetric equilibrium dust cloud into a dust void as the RF power and pressure increase.
  
The present work is devoted to the spontaneously excited converging shock structures observed in a complex plasma medium. Converging shock waves \cite{perry1951production} occur naturally in collapsing spheres and possesses potential application, e.g., shock wave lithotripsy \cite{pearle2012shock}, diamond synthesis \cite{sawaoka2012shock}, and inertial confinement fusion \cite{boehly2011multiple}. All these applications have one thing in common to create extreme conditions, i.e., creating a high concentration of energy at one point. Here, we have observed and analyzed the characteristics of a self-excited converging shock structure in a dusty plasma medium for the first time. A  horizontally confined high-density dust cloud of melamine formaldehyde particles is created in a DC glow discharge plasma. A self-excited compressed dust density pulse has been initiated from the boundary when the discharge voltage is increased. This structure is found to propagate inward towards the center of the dust cloud. We have characterized this self-excited density crest and found that it follows the properties of a shock structure propagating with a velocity greater than the acoustic speed of the medium. The amplitude and width of the shock are measured and found to increase during its propagation. In past studies \cite{tadsen2017amplitude, arora2020excitation}, it has been shown that the amplitude of linear and nonlinear waves increases with the decrease in dust density. However, we have found that the amplitude of the density structure observed in our study follows an opposite trend in the direction of the density gradient. This is because the density structure excited in our study follows the typical characteristics of a converging shock wave.  The strength of the excited shock increases with an increase in discharge voltage. For higher values of discharge voltage, the circular symmetry of the excited structure breaks, and a transition from ring-shaped to blob-like structures is observed. We are contemplating that the excitation of this shock structure is initiated due to the perturbation in the sheath electric field originating when the discharge voltage is increased. We have also performed molecular dynamics simulation (MD) to validate our experimental observations. A density structure is excited in the simulation by applying an electric field in the form of a pulse on the dust cloud at the boundary. The propagation characteristics, such as amplitude and width of the excited circular density structure in simulation, follow a similar trend as observed in the experiments. 

The whole paper is divided as follows. We describe the experimental details and their associated results in Sec. \ref{sec:exp}. The experimental setup and method are presented in Sec.\ref{sec:setup}. The experimental observations and the results are provided in Sec.\ref{sec:exp_results}. In Sec.\ref{sec:sim}, we describe the simulation. The MD simulation details are provided in \ref{sec:sim_paramtr}, and the results associated with simulations are discussed in \ref{sec:sim_results}. Finally, section Sec.\ref{sec:conc} concludes all the experiment and simulation observations.  
 \par
\section{Experiment}\label{sec:exp}
\subsection{Experiment Method}\label{sec:setup}
 \begin{figure}[ht]
\includegraphics[scale=0.39]{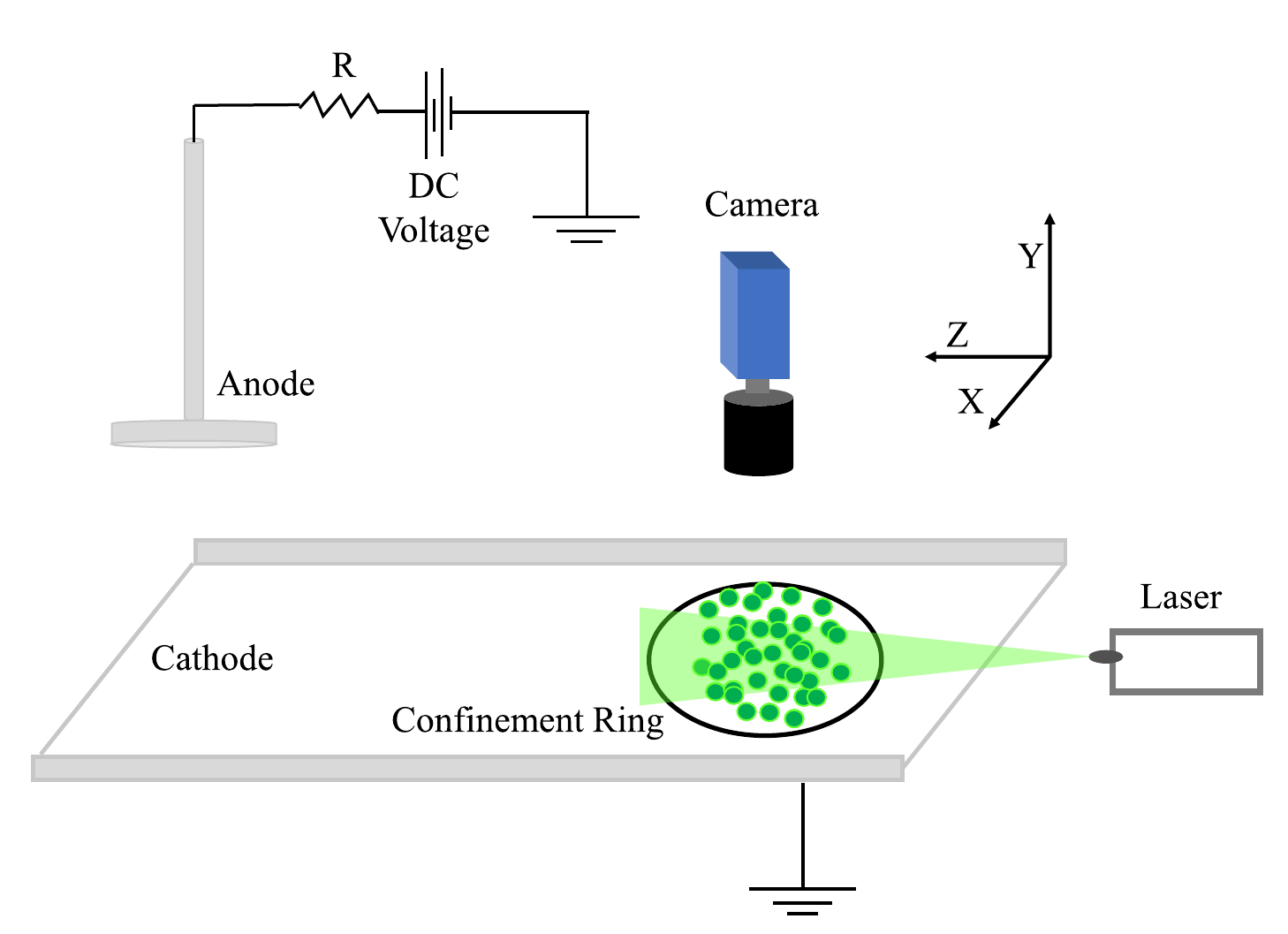}
\caption{\label{fig:setup} Sketch of the experimental set-up of DPEx-II device. A DC power supply generates plasma, and monodispersive particles of diameter 10.66 $\mu$m are suspended to create dusty plasma. Micro-particles are illuminated by a horizontal sheet of laser light and imaged from the top by a high-speed camera.  }
\end{figure}
The experiments related to this study have been performed in a newly built Dusty Plasma Experimental-II (DPEx-II) device. The extensive study on the experimental set-up and plasma diagnostics can be found in Ref. \cite{arumugam2021dpex}. The device has an asymmetric electrode configuration, which has been used to generate DC plasma and large-sized dusty plasma. The schematic of the setup is shown in Fig.\ref{fig:setup}(a). The uneven electrode configuration with a disc shape serves as an anode, and a wide rectangular tray with upward bents serves as a cathode. The DC glow discharge plasma is formed in an argon gas environment with a pressure of $7.0$ Pa and a DC voltage of 320-360 V. The plasma parameters, such as plasma density and electron temperature, are measured with the help of single and double Langmuir probes and are estimated to be 1-6 $\times~10^{15}~\rm /m^3$ and 2-4 eV, respectively.
Mono-dispersive melamine formaldehyde dust particles of diameter 10.66 $\mu$m are introduced into the plasma to create a dusty plasma. The particles levitate above a stainless steel circular ring on the cathode's right end. This ring also serves as a part of the cathode and creates a radial electrostatic trap due to the radial sheath around it. In the plasma environment, the dust particles get negatively charged as the larger number of electrons, being lighter than ions, attach to their surface. These negatively charged dust particles levitate at a certain height from the cathode by the balance of downward gravitational force and upward electric field of the cathode sheath. The radial expansion of the dust cloud is restricted by the radial sheath imposed by the ring called the confinement ring. Using the collision-enhanced model (CEC) \cite{khrapak2006grain}, the dust charge ($Z_de$) is estimated to be $\sim~4.8 \times 10^3~\rm e$. Our estimated charge from CEC theory has also been supported by the experimental verification based on a force balance model in our past study \cite{arumugam2021dpex} using the same experimental device and under similar discharge conditions. In experiments, the levitation of negatively charged dust particles occurs due to the balance of downward gravitational force ($m_dg$) and upward electrostatic force ($QE_z$). The mass density of melamine formaldehyde particle (dust particle) is 1.5 g/cm$^{-3}$, and the radius ($r_d$) is 5.33 $\rm \mu$m which gives the mass of the dust particle to be 9.5$\times~10^{-13}$ kg. The electric field strength measured using an emissive probe in the same device without the presence of dust particles is $E_z\sim$ 2$\times~10^4$ V/m \cite{arumugam2021dpex} for our discharge conditions. The force balance equation $m_dg=QE_z$ estimates the dust charge to be $Q\sim~3\times10^3$e. Although the measurements from the probe are subjected to 20-30 $\%$ error, the estimated charge from the experimentally measured electric field is of the same order as from CEC theory. The 2D dust density is roughly calculated to be $\sim~1\times10^8~/m^2$. The dust particles are illuminated by a horizontally aligned green laser light (XZ plane), and the dynamics are recorded from the top view high-speed camera. 

\subsection{Experimental Results}\label{sec:exp_results}
 \begin{figure*}[ht]
\includegraphics[scale=0.65]{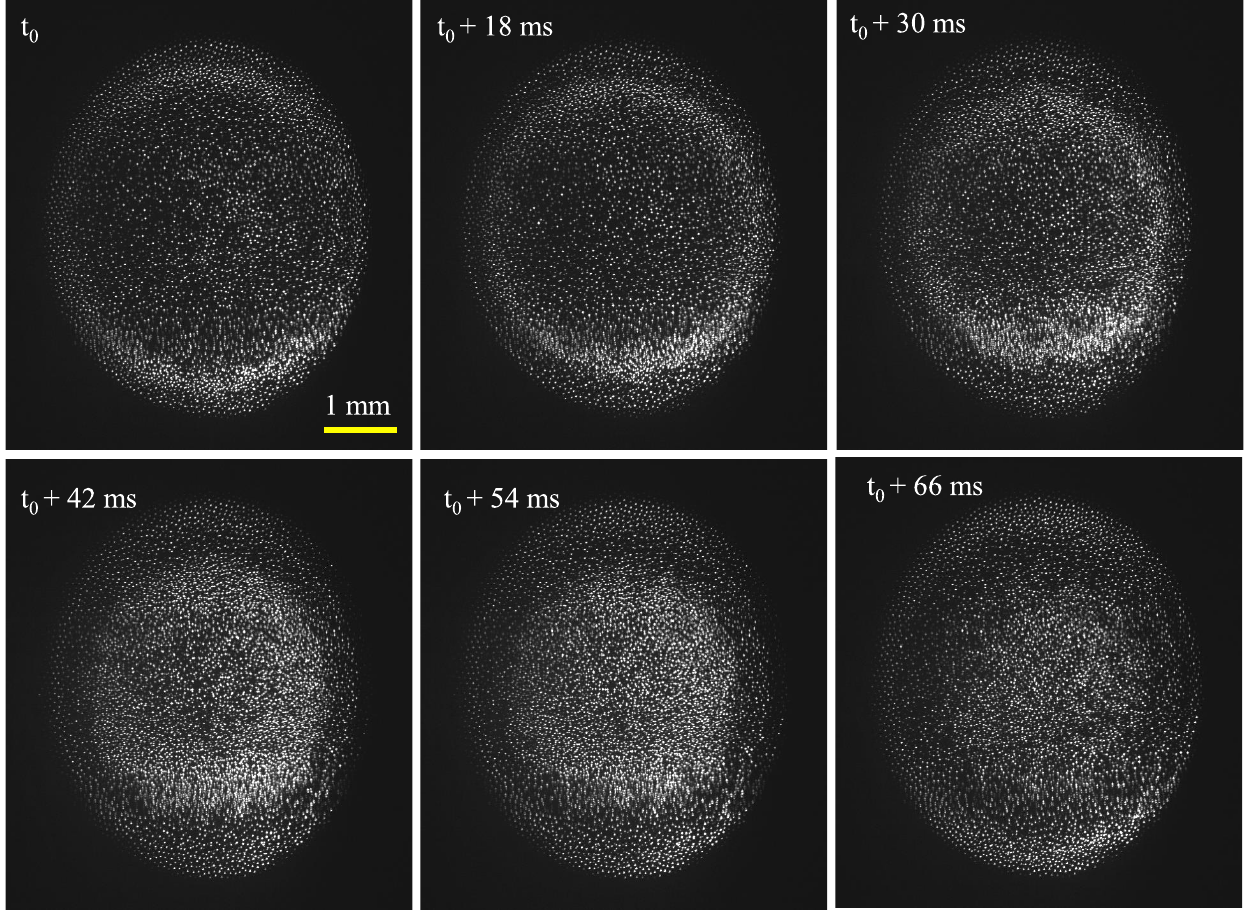}
\caption{\label{fig:image_circular} The images show the snapshots of a dust cloud in which a single-density structure having a circular shape is spontaneously excited in the DPEx-II device. The plasma is formed at $V_d =340$ V and pressure $p=7~Pa$. As time progresses, the density crest propagates radially inward, becoming brighter and wider.}
\end{figure*}
To begin with, an argon pressure of 7 Pa and DC voltage of 320 V is set to create a plasma and dusty plasma. A higher dust density is aimed at introducing more particles in this device than in past experiments. A stable circular dust cloud with high dust density is created at this plasma condition. In this condition, a dust fluid with density $1\times10^8~m^{-3}$ is obtained in a strongly coupled liquid state with an estimated coupling parameter $\Gamma \sim 180$. The coupling parameter $\Gamma =(Q^2/4\pi\epsilon_0ak_BT_d)~exp(-a/\lambda_D)$ is defined as the ratio of inter-particle electrostatic potential energy to the average thermal energy of the particles. Here, $Q,~\epsilon_0,~a,~k_B,~T_d, and ~\lambda_D$ are the particle's charge, the electric permittivity of free space, inter-particle distance, Boltzmann constant, the temperature of the dust particles, and dust Debye length, respectively. The dust Debye length is defined as $\lambda_D=(\epsilon_0k_BT_i/n_ie^2)^{1/2}$, where $T_i,~ n_i$ are the temperature of ions and plasma density, respectively. For our discharge conditions, i.e.,  pressure, $P=7$ Pa, and discharge voltage $V=320$ V, $T_i\sim 0.03~eV$, and  $n_i\sim~1\times~10^{15}~\rm m^{-3}$ \cite{arumugam2021dpex}, we estimated the value of the coupling parameter to be $\Gamma\sim180$. 

In our experiment, the sheath electric field of the horizontal confinement ring has a 3D profile. However, the monodispersive particles mainly form a 2D monolayer levitating at a certain height for our discharge conditions. The vertical component of the electric field, originating from both the cathode and circular ring, is balanced out by the gravitational force acting on the particles. At higher discharge voltages, the particles are compressed in the horizontal plane due to an increase in the horizontal confinement field. As a result, some particles may go down and form a quasi-2D layer. However, a particular monolayer shown in Fig.\ref{fig:image_circular} has the highest populated density and a much larger area in the horizontal plane. As a result, particles levitating in this particular layer are subjected to the stronger sheath electric field of the horizontal confinement ring. As the DC voltage is increased from 320 to 340 V, a single-density compressed structure is excited, propagating in the radially inward direction. The snapshots from the fast camera comprising the dust fluid and excited density compressed structure are shown in Fig.\ref{fig:image_circular}. As time progresses, the density structure propagates radially inward, and the density compression becomes brighter and wider. The last snapshot shows the concentration of energy in the central region where the amplitude and width are higher. It is to be noticed that the luminosity is brighter at the lower edge than the upper edge, as can be seen from Fig.\ref{fig:image_circular}. We are speculating that this non-uniformity comes from the non-uniform perturbation of the sheath electric field, which arises due to the manufacturing imperfection of the metallic ring responsible for the horizontal confinement or due to the inhomogeneous variation of plasma potential or electric field in the radial direction.

 \begin{figure*}[ht]
\includegraphics[scale=0.95]{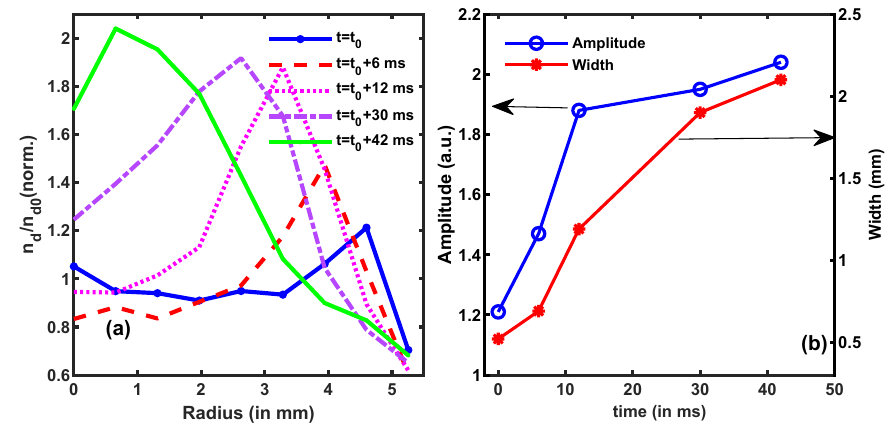}
\caption{\label{fig:amp_width_time} (a) shows the profile of compression factor obtained from different frames of images. (b) shows the temporal profile of amplitude as well as width estimated from the profiles of density crest.   }
\end{figure*}
In order to obtain the quantitative parameters of the excited structure, such as amplitude, width, and velocity, we have estimated the radial density distribution of the dust cloud with respect to its center. The spatial profile of dust density is estimated from the scattered light intensity. The dust density is considered to be proportional to the intensity of light scattered from the dust particles, as the used CCD camera has a linear response to the light intensity. Now, to estimate the one-dimensional (1D) density profile as shown in Fig. \ref{fig:amp_width_time}(a), the whole dust cloud is radially binned by concentric circular strips with a width $dr = 0.1$ mm. The 1D radial density profile is then obtained from the mean intensity of the scattered light within these circular bins. We define $n_d(r)/n_{d0}(r)\sim I(r)/I_0(r)$, where $I_0(r)$ is the intensity at a particular radial distance $r$ when the cloud is stable with no excited density compression, and $I(r)$ represents the same with the excited density structure, as shown in the images of  Fig.\ref{fig:image_circular}. The amplitude of the crest is defined as the peak of $n_d/n_{d0}$, and the width is estimated using the technique mentioned in  Ref. \cite{heinrich2009laboratory} by Heinrich \textit{et. al.}.

Fig.\ref{fig:amp_width_time}(a) shows the evolution of the radial density profile of the dust cloud at different times. The shift of the peak towards the center ($r=0$) with time indicates the inward propagating nature of the self-excited density crest. Fig.\ref{fig:amp_width_time}(b) shows the time evolution of the estimated amplitude and width of the excited density compression. Here, both the amplitude and width of the spontaneously excited density crest increase with time. The obtained amplitude values signify that the structure is nonlinear in nature. The nonlinearity can also be implied from the propagation speed. We have estimated the propagation speed of the density crest by tracking its evolution over time and found it to be $v\sim$ 3 cm/s. We have estimated the dust acoustic speed ($C_d$) using the expression, $C_d=\omega_{pd}\lambda_D$, which has a value $C_d\sim$$1.2$ cm/sec for our discharge conditions. Here, $\omega_{pd} = (Q^2/2\pi\epsilon_0m_da^3)^{1/2} \approx~262 $ Hz and $\lambda_D \approx 45~\rm \mu m $ represent the dust-plasma frequency and dust Debye length, respectively. The estimated Mach number $(M=v/C_d)$ $\sim$ 2.5 indeed shows that the structure is nonlinear in nature. The overall properties of the excited density pulse, e.g., radial shape, nonlinearity in the amplitude, and $M>1.0$, indicate that the excited density crest is nothing but a shock structure. The increase in width during its propagation is another typical characteristic of a shock structure due to the dissipative losses in the medium. Here, in our medium, dissipation can come from strong coupling effects as well as neutrals streaming. However, the increase in amplitude during its journey is interesting despite its propagation along the increasing dust density towards the center of the cloud. Tadsen \textit{et. al.} \cite{tadsen2017amplitude} experimentally showed the propagation characteristics of a linear DAW in an inhomogeneous dust cloud and found that the amplitude of linear waves increases with a decrease in dust density. Recently, Arora \textit{et. al.} \cite{arora2020excitation} excited the shock waves and studied the propagation characteristics in a linear gradient of dust density. They found that the shock's amplitude and width increase with a decrease in dust density. In the present study, the increase in the amplitude of the excited shock propagating towards the increasing dust density can be understood as follows. The shock waves have a high energy density distributed in a thin streak, and its converging nature signifies that the area is reducing; hence, energy density increases due to the conservation of energy followed by an acceleration of the shock wave. Thus, in our case, the rise in amplitude of the shock structure is attributed to its converging nature.

 \begin{figure}[ht]
\includegraphics[scale=0.65]{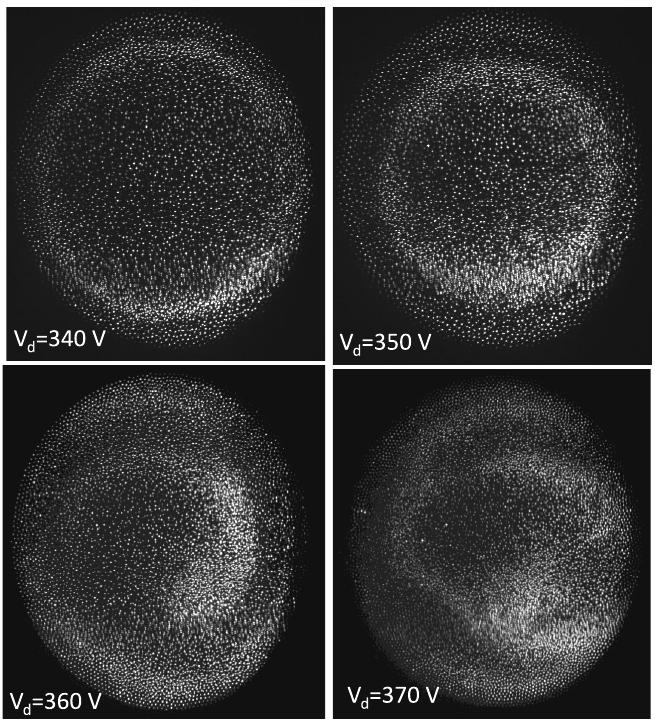}
\caption{\label{fig:image_voltage} Snapshots show the excited density crest in a dust cloud at t= 18 ms for four different discharge voltages. The background pressure is  kept constant at 7 Pa  }
\end{figure}

As the shock structure in our medium is spontaneously excited by increasing the discharge voltage, we have performed a series of experiments by changing the discharge voltage. Fig.\ref{fig:image_voltage} shows the snapshots of dust clouds with the excited shock structure for four different discharge voltages at a constant time. Here also, an excitation of shock structure is seen at a discharge voltage of 350 V, similar to 340 V. The only difference is the position of the crest and compression factor. The structure travels more inward over a certain time at a higher voltage. This implies that the propagation speed increases with an increase in discharge voltage. However, if we keep increasing the discharge voltage, the circular symmetry breaks into segments. This is attributed to the non-uniformity in the amplitude of excited density crest at different azimuthal positions, as mentioned in the discussion related to Fig.\ref{fig:image_circular}. At higher discharge voltages, the sheath electric field and, consequently, the perturbation strength increases. Thus, the scale of the non-uniformity over different azimuthal positions in the amplitude of the excited structure increases at higher voltages. The velocity of the excited structure is proportional to its amplitude. Thus, the non-uniformly excited structure will move with different velocities at different azimuthal locations, resulting in the breakup of the ring shape, forming blob-like structures. 

 \begin{figure}[ht]
\includegraphics[scale=0.65]{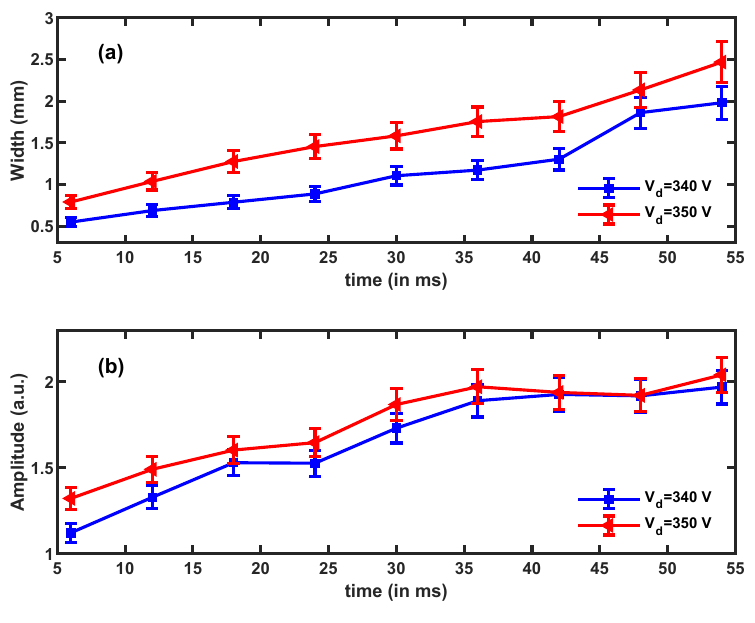}
\caption{\label{fig:amp_width_twov_exp} (a) Width (b) Amplitude of excited circular structure as a function of time for two different voltage.    }
\end{figure}
 \begin{figure}[ht]
\includegraphics[scale=1]{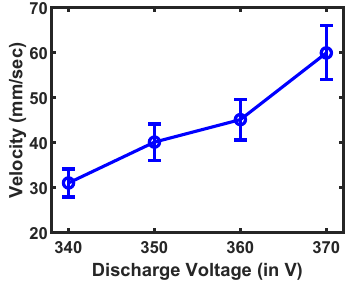}
\caption{\label{fig:vel_exp} Variation of velocity with discharge voltage estimated by tracking a density crest over time.    }
\end{figure}

 The time evolution of the width and amplitude of the shock for two different discharge voltages are shown in Fig. \ref{fig:amp_width_twov_exp}. The subplots show that the amplitude and width of the shock increase as time progresses for each discharge voltage. However, at a particular instant of time, both the amplitude and width of the shock have higher values for higher discharge voltage. Since the exact cause of the excitation of the shock structure observed in our study is still unclear, we hypothesize that the abrupt change in the sheath electric field, triggered by the increasing discharge voltage, could potentially be the reason for the spontaneously excited perturbation. As the outermost particles within the dust cloud predominantly encounter the horizontal sheath electric field, the density perturbation is excited around the periphery of the dust cloud. With the increase in discharge voltage, the electric field strengthens, and the perturbation in the electric field increases. Hence, the density compression of the excited structure increases at higher voltages. We have also estimated the velocity of the inward propagating shock structure for different discharge voltages, as shown in Fig.\ref{fig:vel_exp}. It has been found that the structure excited in higher discharge voltage moves with higher velocity.

\section{Molecular Dynamics Simulation}\label{sec:sim}

In order to understand and qualitatively explain the various features of our experimental findings, a set of three-dimensional (3D) molecular dynamics (MD) simulations has been performed. We suspect that the spontaneous fluctuation in the radially inward electric force originating from the sheath electric field of the confinement ring is responsible for the excitation of the inward propagating dust density crest observed in our experiment. We aim to model this phenomenon qualitatively using MD simulations. In our simulation, a monolayer fluid of the charged dust particles is created using the experimental parameters, and a density crest is excited in this dust fluid by imposing an external perturbation in the radial confinement electric field. The details of the simulation and corresponding results are described in the following subsections.

\subsection{Simulation Details}\label{sec:sim_paramtr}

We have used an open-source massively parallel classical MD code LAMMPS \cite{plimpton1995fast} to perform three-dimensional MD simulations. Initially, two thousand point particles with charge $Q = 4.8\times 10^3e$ and mass $m_d = 9.5\times 10^{-13}$ kg are randomly distributed inside a 3D simulation box with length $L = 0.05$ cm in all three directions. Here, $e$ represents the charge of an electron. Periodic boundary conditions in all three directions have been considered. Particles are assumed to interact via Yukawa pair potential. The electric field in the form of $\mathbf{E}_{x}(x) = -(m_d\omega_r^2/Q)(x-L/2)\hat{x}$ and $\mathbf{E}_{y}(y) = -(m_d\omega_r^2/Q)(y-L/2)\hat{y}$ has been applied in the x and y-directions, respectively, providing parabolic confinement of particles in the horizontal plane. In the vertical direction, we have considered the force due to gravity $m_dg(-\hat{z})$ (i.e., acting vertically downward) and force associated with an externally applied electric field $\mathbf{E}_z(z) = QE_{z0}\exp{(-\alpha z})\hat{z}$ (acting vertically upward for the negatively charged particle), mimicking the cathode sheath electric field in experiment \cite{sickafoose2002experimental,arumugam2021dpex}. A combined effect of these two forces provides an effective parabolic confinement of particles in the vertical direction \cite{maity2019molecular}. In our simulations, the values of $\omega_r$ and $\alpha$ are considered to be $1.0$ Hz and $500$ m$^{-1}$, respectively. The equation of motion of any $i$th particle can be expressed as,

\begin{equation}
 m_d\frac{d^2\mathbf{r}_i}{dt^2}= -Q\sum_{j=1}^{N-1}\nabla U(r_{i,j}) -m_d\mathbf{g} + Q(\mathbf{E}_{x} + \mathbf{E}_{y} + \mathbf{E}_{z}),  
\end{equation}

where $r_i$ and $r_j$ are the positions of the $i$th and $j$th
particles at a particular time, respectively, and $U(r_{i,j})=(Q/4\pi\epsilon_0|\bf{r}_j-\mathbf{r}_i|)\exp{(-|\mathbf{r}_j-\mathbf{r}_i|/\lambda_D)}$ represents Yukawa pair potential between $i$th and $j$th particle. Here, $N$ represents the total number of particles.

Initially, particles were relaxed to a thermal equilibrium state with a desired temperature in the presence of a Nose-Hoover thermostat \cite{nose1984molecular, hoover1985canonical}. Later, the thermostat was disconnected, and the system was allowed to evolve under a micro-canonical ensemble where the system's total energy remained conserved. In equilibrium, under the influence of all the confining forces, particles levitate at a certain height (i.e., at $z = L/2$ for our chosen values of $E_{z0} = (m_dg/Q)\exp{(\alpha L/2)}$), forming a monolayer disc in the horizontal (x-y) plane with radius $R$ with respect to the center of the dust cloud. Now, we introduce a perturbation to the system by applying electric fields in the following forms,

\begin{equation}
    \mathbf{E}_{xp} = E\delta(t-t_0)\frac{(x-L/2)}{|(x-L/2)|}\hat{x},
\end{equation}

\begin{equation}
    \mathbf{E}_{yp} = E\delta(t-t_0)\frac{(y-L/2)}{|(x-L/2)|}\hat{y},
\end{equation}
 
where $t_0$ is the time of the perturbation and $\delta$ represents the Dirac’s delta function. The electric perturbation has been applied only to the particles for which the radial locations with respect to the center of the monolayer, $r_i>0.9 R$, i.e., particles residing at the outer radial locations. This results in an electrostatic force acting on the outer particles of the monolayer along the radially inward direction. As mentioned earlier, MD simulations have been performed only to qualitatively model our experimental observations concerning the shock characteristics, i.e., amplitude, width, velocity, and its converging nature. In simulations, the amplitude of the electric field perturbation is taken to be an arbitrary value so that the initial density perturbation (i.e., amplitude and profile) triggered in the monolayer remains in the same order as observed in the experiment. We have characterized the evolution of dust fluid under the influence of this electrostatic perturbation and presented them in the following subsection. 

\subsection{Simulation Results}\label{sec:sim_results}
 \begin{figure}[ht]
\includegraphics[height = 6.7cm,width = 8.0cm]{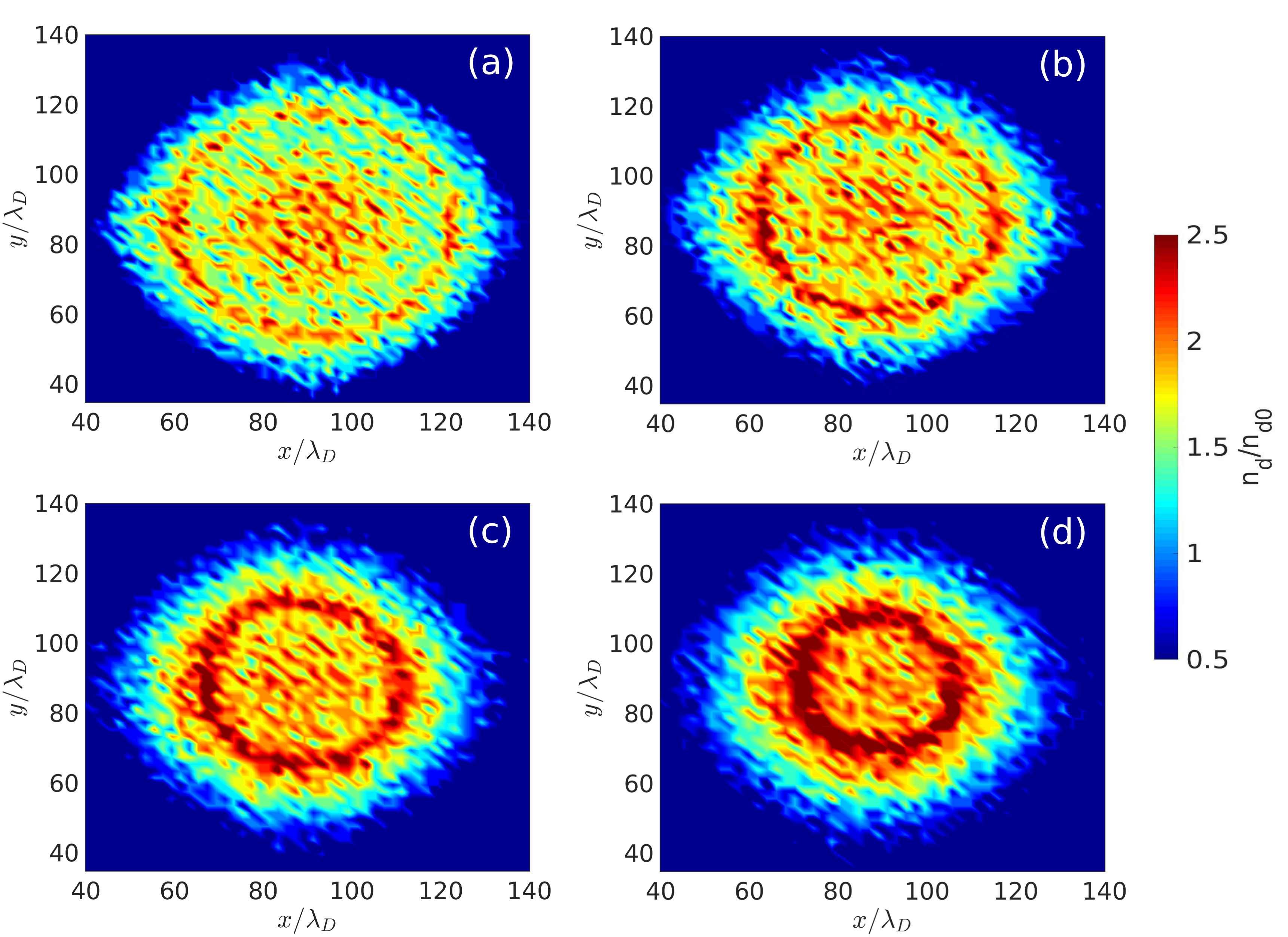}
\caption{\label{fig:density_pcolour_subplot}  Distribution of the particle number density ($n_d$) in the x-y plane of the monolayer at time (a) $\omega_{pd}t = 8$, (b) $\omega_{pd}t = 10$, (c) $\omega_{pd}t = 12$, and (d) $\omega_{pd}t = 14$. In this case, the electric field perturbation is considered to be $E/E_0 = 0.8$.}
\end{figure}
 \begin{figure}[ht]
\includegraphics[height = 5cm,width = 6.0cm]{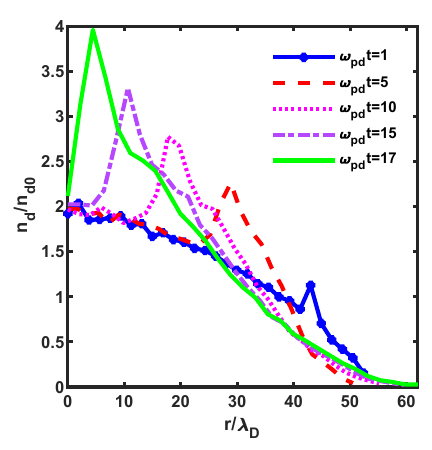}
\caption{\label{fig:time_evol_s}  Radial density profile of the monolayer at different times of the simulation run with an initial electric field perturbation $E/E_0 = 1.1$. It is clearly seen that a density hump with a distinct peak appears as a result of the initial electric field perturbation, and it moves radially inward with time.}
\end{figure}
\begin{figure}[ht]
\includegraphics[scale=0.6]{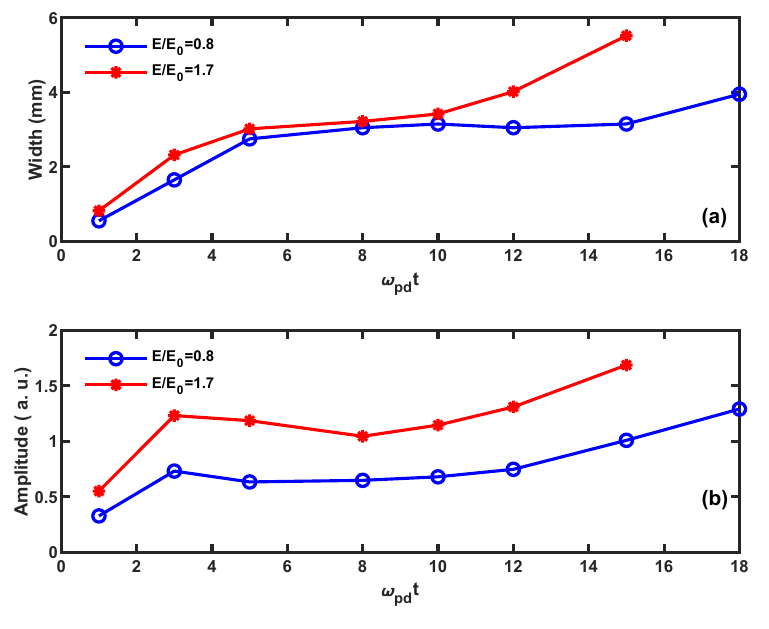}
\caption{\label{fig:amp_width_twov_s} The width and amplitude of the density hump as a function of time have been sown in subplots (a) and (b), respectively, for two different cases with initial perturbation $E/E_0 = 0.8$ ( blue circles) and $E/E_0 = 1.7$ (red asterisk).}
\end{figure}

In equilibrium (without external perturbation), the dust cloud has a radially inhomogeneous density profile with a higher density near its center. The perturbation in the horizontal electric fields in the form of a pulse creates a density hump around the outer boundary of the 2D monolayer. Fig.\ref{fig:density_pcolour_subplot} (a)-(d) shows the distribution of particle number density in the x-y plane of the monolayer at times $\omega_pdt=8,~10,~12,~14$, respectively, for a particular case when the perturbation in the electric field is $E/E_0=0.8$. Here, $E_0 = Q/4\pi\epsilon_0a^2 \sim 193$ V/m is the average inter-particle electric field where $a = 1/(n_{d0}\pi)^{1/2}$ represents the average inter-particle distance and $n_{d0} = N/\pi R^2$ is the average $2D$ density of the monolayer. The subplots of Fig.\ref{fig:density_pcolour_subplot} show that a single-density crest (red color) is created in the dust fluid in response to the electric perturbation. As time progresses, the density compression moves radially inward and becomes wider, similar to the observation in experiments. We have also estimated the radial density profile of the dust cloud under the influence of the external perturbation at different simulation times and presented them in Fig.\ref{fig:time_evol_s}. The figure shows that the peak of density, representing the density crest, moves towards the center, ensuring the radially inward propagation of the crest. We have estimated the velocity of this structure and found that it propagates with a supersonic speed. It is also seen that the density structure has a triangular shape and its peak value increases during its propagation. Thus, the properties of the excited density structure, e.g., its shape, speed, and nonlinearity in the amplitude, ensure that it follows the same characteristics of a converging shock. We have extracted the quantitative parameters, such as amplitude and width, from the radial density profile and shown in Fig.\ref{fig:amp_width_twov_s}. 
 \begin{figure}[ht]
\includegraphics[scale=1]{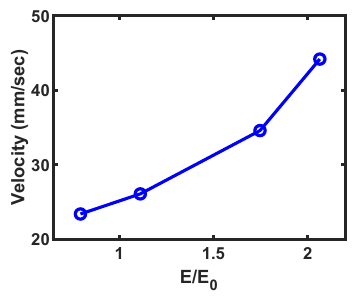}
\caption{\label{fig:vel_s}  Velocity of the radially inward propagating density crest for different values of initial perturbation ($E$) has been shown.}
\end{figure}

In Fig.\ref{fig:amp_width_twov_s} (a) and (b), we have shown the time evolution of width as well as amplitude for two different electric field perturbations. The amplitude is defined as the maximum of $n_d/n_{d0}$, and the width is determined using the technique mentioned in \cite{heinrich2009laboratory}. The figure clearly shows that as time evolves, the amplitude of the excited density crest increases, and it becomes wider. Here, the width of the shock structure increases because of dispersion originating from the strong coupling effect of the medium. At the same time, the amplitude increase in its inward propagation is attributed to its converging nature and can be understood from the energy conservation point of view. As mentioned in section \ref{sec:exp_results}, in our experiments, we increased the discharge voltage, which increased the sheath electric field, and found that the density crest is excited with a larger amplitude and width. In the simulation, we have replicated the phenomena by increasing the amplitude of electric field perturbation. Here also, at a given time, the amplitude and width of the excited shock pulse have higher values as compared to lower perturbation, and its amplitude increases as time progresses, as observed in our experiments. 

The velocity variation estimated from the simulation runs with the increased electric perturbation has been shown in Fig.\ref{fig:vel_s}. The figure depicts that velocity increases as we increase the amplitude of perturbation akin to Fig.\ref{fig:vel_exp} obtained in the experiments. This can be explained as follows. As we increase the amplitude of perturbation, the amplitude of the density hump created at the boundary increases. A typical characteristic of a shock is that its propagation speed increases with an increase in nonlinearity. Since nonlinearity depends upon the amplitude of the excited structure, a density crest with higher amplitude propagates with higher velocity. Henceforth, all the results obtained in the simulation qualitatively explain the experimental observations.


\section{Summary}\label{sec:conc}

This study investigates the evolution of a self-excited converging shock observed in a complex plasma medium. A nonlinear density structure has been excited in a complex plasma experiment performed in the DPEx-II device upon increasing the discharge voltage after a threshold. The structure was first initiated around the outer boundary of a stable dusty cloud and found to propagate radially inward. The properties of this structure have been analyzed, and it is found that it follows the characteristics of a converging shock. Contrary to the previous reports, our study reveals that the shock amplitude increases despite moving toward higher density. It has been understood that this phenomenon is a consequence of the converging nature of the shock structure. We have also experimentally analyzed the dependence of discharge voltage on the properties of this self-exited shock structure. It has been found that the shock's amplitude and width at any particular time of evolution have higher values for higher discharge voltages. In our study, we have also performed three-dimensional MD simulations to compare our experimental findings qualitatively. We have shown that our simulation results are in good agreement with the experimental observations, providing a qualitative explanation of the observed phenomena.

\section{Acknowledgment} 

The authors would like to acknowledge Dr. Pintu Bandyopadhyay and the Institute for Plasma Research, India, for providing the experimental facility to perform this study. The authors would also like to thank Dr. Saravanan Arumugam for the technical help during the experiments. The authors also wish to thank the anonymous referees, whose comments and suggestions have significantly improved the manuscript.

\bibliography{reference}
\end{document}